# DIAMOND CRYSTAL WITH Y-DEFECTS: SPECTROSCOPY AND TRANSMISSION ELECTRON MICROSCOPY

**A.A. Shiryaev[1], E.A. Vasilev[2], A.L. Vasil'ev[3,4], V.V. Artemov[3], N.V. Gubanov[5], D.A. Zedgenizov[5]**

[1]*Frumkin Institute of physical chemistry and electrochemistry RAS, 119071, Moscow, Leninsky pr. 31b4, Russia*

[2]*Saint Petersburg mining university, Saint Petersburg, Russia*

[3] *Shubnikov Institute of Crystallography, Kurchatov Complex of Crystallography and Photonics, National Research Centre "Kurchatov Institute," Moscow, 119333 Russia*

[4]*Moscow Institute of Physics and Technology, 9 Institutskiy Per., Dolgoprudny, Moscow Region, 141701, Russian Federation*

[5]*Zavaritsky Institute of Geology and Geochemistry UB RAS, Ekaterinburg, Russia*

The paper presents results of investigation of a natural Ib-IaA diamond containing Y-defects from Yubileinaya kimberlite pipe, Yakutia. Analysis of spatial distribution of nitrogen-related A and C centers and intensity of IR absorption at Raman frequency (1332 $cm^{-1}$) reveals anticorrelation between these defects. Transmission electron microscopy of a zone with abundant Y-defects shows presence of dislocations in various configurations and numerous clusters of point defects generated by non-conservative dislocation movement. Extended defects with shape resembling thin (1-3 nm) rhombic plates with the largest dimension up to 5-20 nm are observed. Analysis of contrast of these defects shows that they represent nanosised voids (vacancy clusters). It is suggested that the defects were formed by annihilation of dislocation dipoles with subsequent growth by consumption of vacancies produced by non-conservative motion of dislocations. Upon excitation by 787 nm laser, numerous narrow photoluminescence lines are observed between 800-900 nm, their intensity and position show pronounced spatial heterogeneity on scale of few microns. Qualitatively similar behavior of photoluminescence was earlier noted for hydrogenated nanodiamonds. It is suggested that unusual behavior of the luminescence lines may be explained by recombination processes on internal walls of the discovered nanovoids.

## Introduction.

Infra-red (IR) spectra of natural diamonds with low nitrogen aggregation state (type Ib-IaA) often show unusual absorption bands in one-phonon region. Although this fact was established long time ago [Clark and Davey 1984; Collins and Mohammed, 1982; Woods and Collins, 1983], such diamonds remain poorly studied. Spectra of many type Ib-IaA diamonds contain an unusual band with complex shape characterized by a broad maximum at ~1145-1150 $cm^{-1}$, less intense features in the range 1220-1270 $cm^{-1}$ and set of weak sharp peaks between 1340-1400 $cm^{-1}$ [Hainschwang et al., 2012]. In [Reutsky et al., 2017] a positive correlation between the main maximum at 1150 $cm^{-1}$ and a IR absorption peak at diamond Raman frequency (1332 $cm^{-1}$) was established. This absorption system is assigned to a so-called Y-center; its model is not elucidated yet. This absorption was not reported for synthetic diamonds.

In pioneering study describing the Y-center [Hainschwang et al., 2012] it was proposed that it represents a novel mode of occurrence of single substitutional nitrogen atom ($N_s$) in diamond structure. Based on experimental study of natural diamonds it was suggested that the Y-center consists of N-C-N complex (an analogue of the N1 defect) and does not comprise interstitials [Nadolinny et al., 2025]. From the analysis of behavior of the Y-center and of nitrogen defects [Day et al., 2025a] propose that the Y-center is related to interstitials and may represent complexes such as ($N_s^0$-$N_i$) or ($N_s^0$-$C_i$). Note, however, that this hypothesis is based on rather strong assumptions about nature and strength of IR absorption of these defects; validity of these assumptions is, unfortunately, not proven. Besides absorptions in IR, numerous luminescing defects are observed in such diamonds [Hainschwang, 2014; Hainschwang et al., 2012; Kupriyanov et al., 2020]. Behavior of some photoluminescence (PL) systems and of the Y-defects during laboratory annealing imply possible correlation between the Y-defects, IR absorption band at 1060 $cm^{-1}$ and PL band peaked at 690 nm. According to some studies, the IR band at 1060 $cm^{-1}$ is related to the oxygen impurity in diamond lattice [Malogolovets and Nikityuk, 1981; Palyanov et al., 2016]. Association of the Y-defects with oxygen was also inferred from their presence in diamonds showing prominent absorption at 480 nm [Hainschwang, 2014; Hainschwang et al., 2020], possibly induced by the oxygen impurity [Gali et al., 2001]. However, the diamonds showing the 480 nm band also contain Ni-related systems [Lai et al., 2024]. Therefore, at present no universally accepted model of the Y-center is available.

In this contribution, we present results of investigation of a natural diamond with the Y-centers employing IR absorption spectroscopy, photo- (PL) and cathodoluminescence (CL) and Transmission Electron Microscopy (TEM).

## Samples and methods.

After IR screening of type Ib-IaA diamonds from Yubileinaya kimberlite pipe (Yakutia) a diamond crystal with cubic habit and strongest absorption by Y-defects was selected. IR absorption spectra were recorded at VERTEX-70 FTIR spectrometer equipped with Hyperion1000 microscope. The spectra were recorded in the 400-7800 $cm^{-1}$ range at 1 $cm^{-1}$ resolution; 32 scans per point; aperture 100×100 μm. In addition, IR mapping was performed. The spectra were normalized to diamond lattice absorption in two-phonon region.

The nitrogen concentration in the A and C defects was calculated using calibrations from [Boyd et al., 1994; Lawson et al., 1998] as following: $N_A = 16.5 \times \alpha(A_{1282})$, ppm, $N_C = 37 \times \alpha(C_{1344})$, ppm, where $\alpha(A_{1282})$ and $\alpha(C_{1344})$ – absorption coefficients of corresponding defects at 1282 and 1344 $cm^{-1}$. The one-phonon region comprises overlapping contributions of C, A, Y defects, carbonate and silicate microinclusions. Decomposition of the spectral envelope was performed using shapes of individual bands using SpectrExamination software [Kovalchuk, 2024], and Excel macro Caxbd_Inherit_2024_Ib [Day et al., 2025a].

Photoluminescence (PL) spectra were recorded using Renishaw InVia spectrometer with ×50 objective at 77 K in a Linkam THMS600 cryostat. For excitation $Ar^+$ laser with $\lambda_{ex}$ = 488 nm (490-750 nm range) and semiconductor laser with $\lambda_{ex}$ = 787 nm (790 - 1050 nm) were employed. No correction for wavelength-dependent detector sensitivity was applied. Optical images were recorded using Leica M205 stereomicroscope. Cathodoluminescence (CL) images were obtained with a "cold" cathode (plasma excitation) at 5 kV accelerating voltage and current of 100-200 μA; CL spectra at room temperature were recorded with MayaPro 2000 spectrometer.

Using $Ga^+$ ion beam three foils with orientations close to (110) and (111) were cut in zones with markedly different intensity of IR absorption by the Y-defect. The foils were studied at a transmission electron microscope OSIRIS TEM/STEM equipped with a high angle annular dark field (HAADF) and energy-dispersive detectors at 200 kV voltage.

## Experimental results.

### *Infra-red and photoluminescence spectroscopy*

Optical and luminescence photographs of the initial crystal and of a flat polished plate made from it are shown in Fig. 1. In IR absorption spectra (Fig. 2a,b) bands due to the A- (maximum at 1282 $cm^{-1}$), C- (1344 $cm^{-1}$), and Y-defects are observed. Absorption at diamond Raman frequency (1332 $cm^{-1}$) is definitely partly related to the Y-defects; however, contribution of $N^+$ (the X-center) or other, yet unidentified defects is not excluded (see Section Discussion). The 3107 $cm^{-1}$ due to nitrogen-hydrogen defect is weak or absent. At the same time, absorption peaks typical for cuboids with Y-centers are observed in the 2800-3300 $cm^{-1}$ range. The sample is zoned; relative intensities of the defects vary between the zones (Fig. 2 c-f). In spectra of individual zones inclusions'-related bands are observed: $v_2$, $v_3$ bands of carbonate anions (877 and 1450 $cm^{-1}$), $H_2O$ bending vibrations (1600-1700 $cm^{-1}$), complex set of broad bands between 1000-1300 $cm^{-1}$ (possibly silicates) and a band at ~3680 $cm^{-1}$, likely related to OH-groups in mica (e.g., in phlogopite). In a zone with the highest absorption by the Y-defects, strength of the 1332 $cm^{-1}$ band reaches 2.7 $cm^{-1}$. The concentration of the A- and C-defects in this zone is minimal; the absorption at 1344 $cm^{-1}$ is less than 0.3 $cm^{-1}$ (Fig. 2c). At the crystal periphery, the concentration of the A- and C-defects reaches 70 and 166 ppm, respectively, whereas the absorption at 1332 $cm^{-1}$ drops to 1.4 $cm^{-1}$. Note that reliable estimation of the A-defect concentration is possible only at the crystal periphery; in the central zone the spectral envelope in the one-phonon region is too complex to allow reliable extraction of the A defect contribution at 1282 $cm^{-1}$.

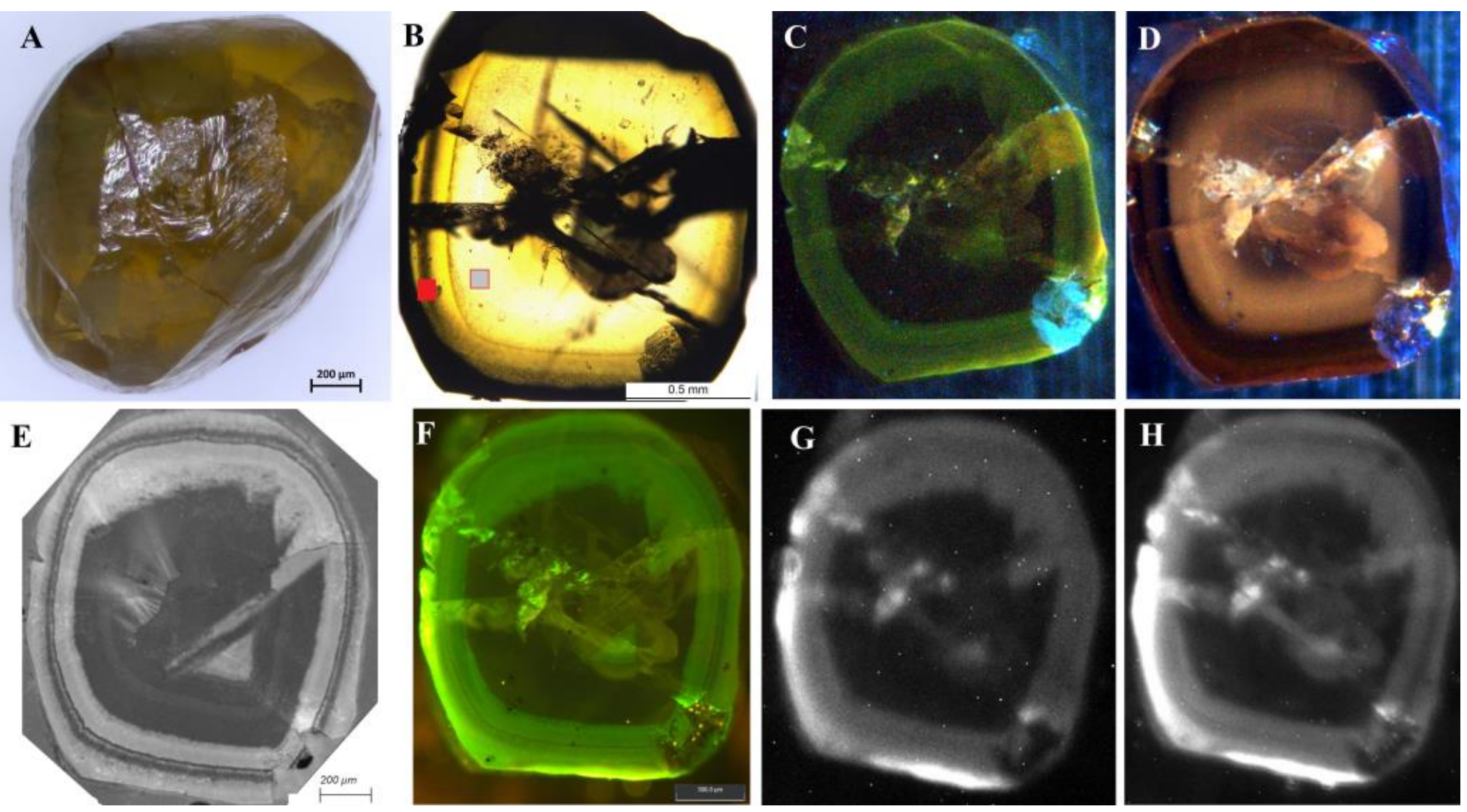

Fig. 1. Optical, photo- and cathodoluminescence images of the U-14 crystal (A) and of the plate cut from it. A – optical photograph of the whole crystal; B – the plate in transmitted light. Squares mark regions from which the foils for the TEM investigation were extracted. C, D – photoluminescence images at excitation at 220 nm (C) and 360 nm (D). E-H – cathodoluminescence images (E – excitation with electron beam; other images – with plasma). E, F – black-and-white and polychromatic images. G, H – images from a multi-spectral camera. G – 500±8 nm channel, H - 550±10 nm channel.

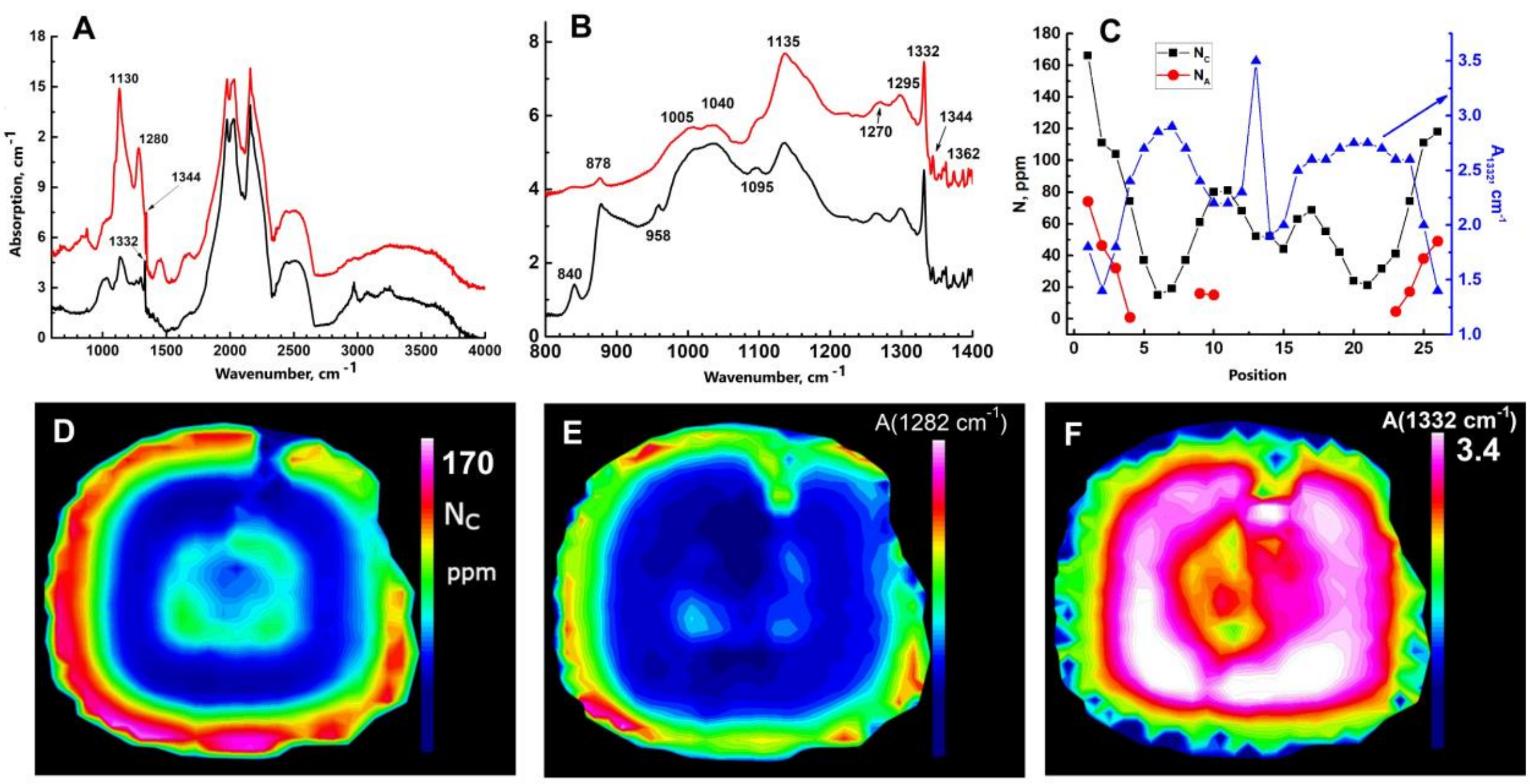

Fig. 2. IR spectroscopy. A – IR absorption spectra of the U-14 diamond in zones with high content of A- and C-defects (red) and Y-centers (black). The spectra are shifted vertically for clarity. B –one-phonon region in different zones. C – Concentration profiles of the A and C-defects and of absorption at 1332 cm$^{-1}$; the profile passes through the crystal center. D-F – maps of A- and C- nitrogen-related defects and of absorption at 1332 cm$^{-1}$. E shows map of intensity at 1282 cm$^{-1}$, see text for detail.

The PL spectra excited at 488 nm (Fig. 3) show the H3 system, a system with zero-phonon line at 525 nm, broad structureless band peaked at ~580 nm, narrow lines 689 and 698 nm. Lines at 630, 632, 638, 685 nm are superposed on the structureless band. The relative intensities of the PL systems vary between the zones. In the zone with the highest Y-defect absorption lines at 622, 624, 630, 632 nm are present. There is no qualitative difference between the center and periphery of the crystal.

In the spectrum excited at 787 nm a very intense line at 904 nm with satellites at 930 and 940 nm is observed; in the peripheral part the H2 (986 nm) is stronger (Fig. 3). More than 40 narrow lines are observed between 800-880 nm. At room temperature individual lines are not resolved and are strongly broadened. Decrease of the relative laser power from 100 to 1% does not induce changes in the spectra. A peculiar feature of these lines is irregular variations of relative intensity and positions of these lines on spatial scale of few micorns: the inset in Fig. 3 shows spectra recorded at very small (~1 μm) distance from each other. No similar features are observed with excitation at 488 nm in the range below the diamond Raman line (521.9 nm). A possible explanation of the PL “blinking” is proposed in the Discussion section.

Cathodoluminescence spectra of the central and peripheral parts are qualitatively similar; however, the latter is notably brighter. Large width of the CL bands precludes unambiguous assignment, but large contributions of the H3 defect and of broad A band seem plausible.

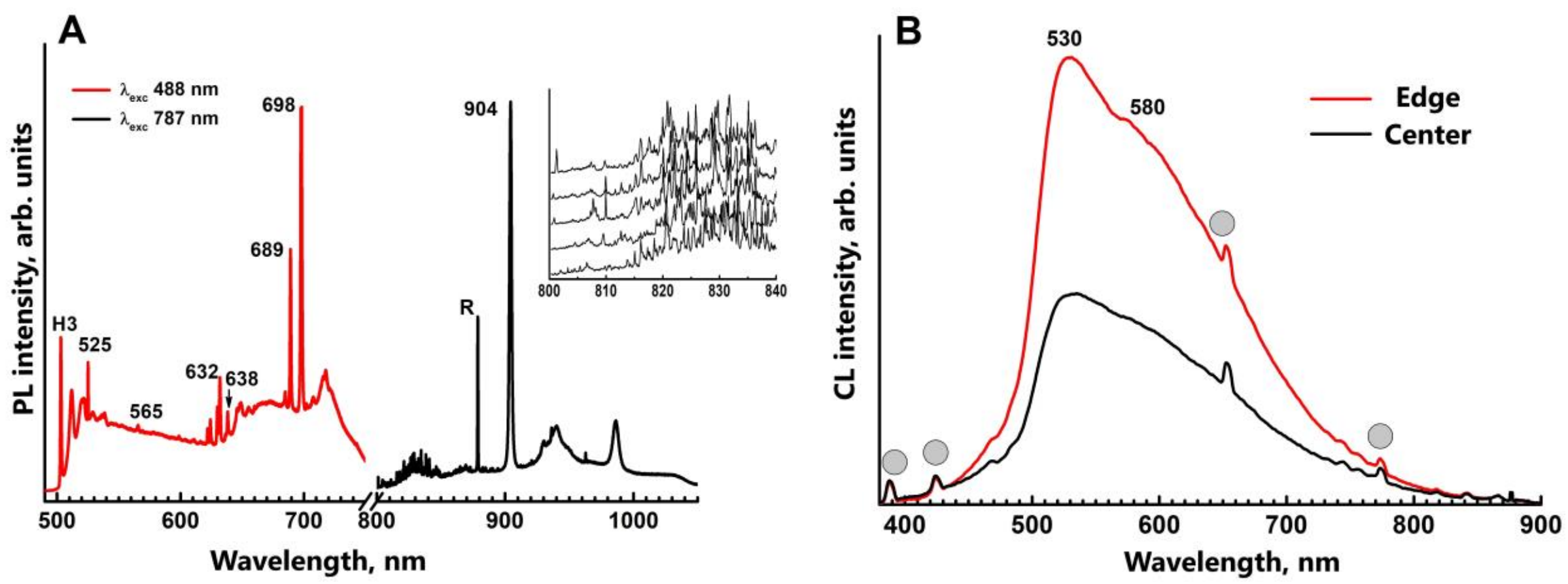

Fig. 3. Luminescence spectra of diamond U-14. A – photoluminescence of a zone rich in Y-defects at excitation ($\lambda_{ex}$) at 488 and 787 nm at 77 K. Inset shows spectra recorded at spots ~1 μm from each other. R – Raman line. B – Cathodoluminescence spectra of central and peripheral parts of the crystal. Gray symbols – background plasma.

*Transmission electron microscopy.*

Transmission electron microscopy reveals presence of three main types of defects: 1) polyphase inclusions with sizes between tens and 300 nm; 2) numerous dislocations in various configurations; 3) strongly flattened nanosized defects with shapes qualitatively resembling rhombic plates.

*Inclusions.*

Although study of polyphase mineral and fluid inclusions is beyond scope of the current work, obtained information is summarized below. In the intermediate zone of the crystal (rich in Y-defects) several cavities with clear crystallographic faceting and filled with mineral phases are observed (Fig. 4,5). In some cavities, the minerals are still present, even despite possible opening during sample preparation. Maps of elemental distribution suggest that one of the minerals was a Ca-Fe-Mg carbonate; presence of KCl and of a complex aluminosilicate (phlogopite?) is also plausible. The presence of carbonates is consistent with IR spectra. Some of the vertices of the inclusions generate dislocations and, possibly, microcracks.

Contrast of the images may imply that that at least one of the cavity walls may be stepped rather than flat. However, the contrast may be explained not only by the индукционной штриховке, since moire pattern may be partly responsible for the observations and testify that the inclusions are crystalline.

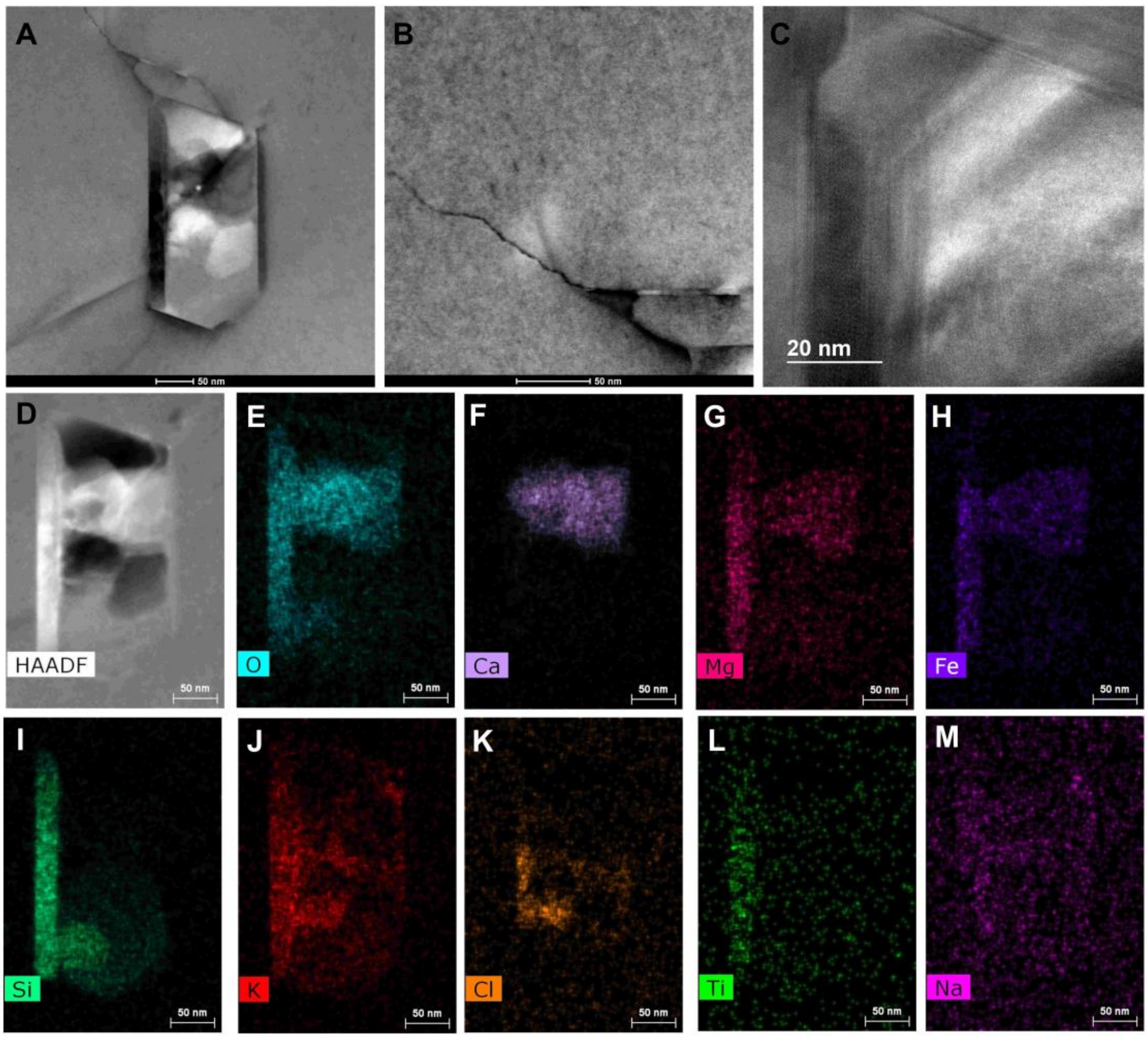

Fig. 4. Polyphase inclusion in U-14 diamond. A – bright field image. B – dislocations or a microcrack at a vertice of the inclusion. A flattened defect parallel to a (111) face of the inclusion is well visible, see text for detail. C – enlarged image of a part of the inclusion. The contrast indicate crystallinity of the inclusion (moire pattern due to superposition of two lattices); it is hatching of the cavity wall is also possible. D – HAADF image of the inclusion; E-M – elemental maps of the inclusion.

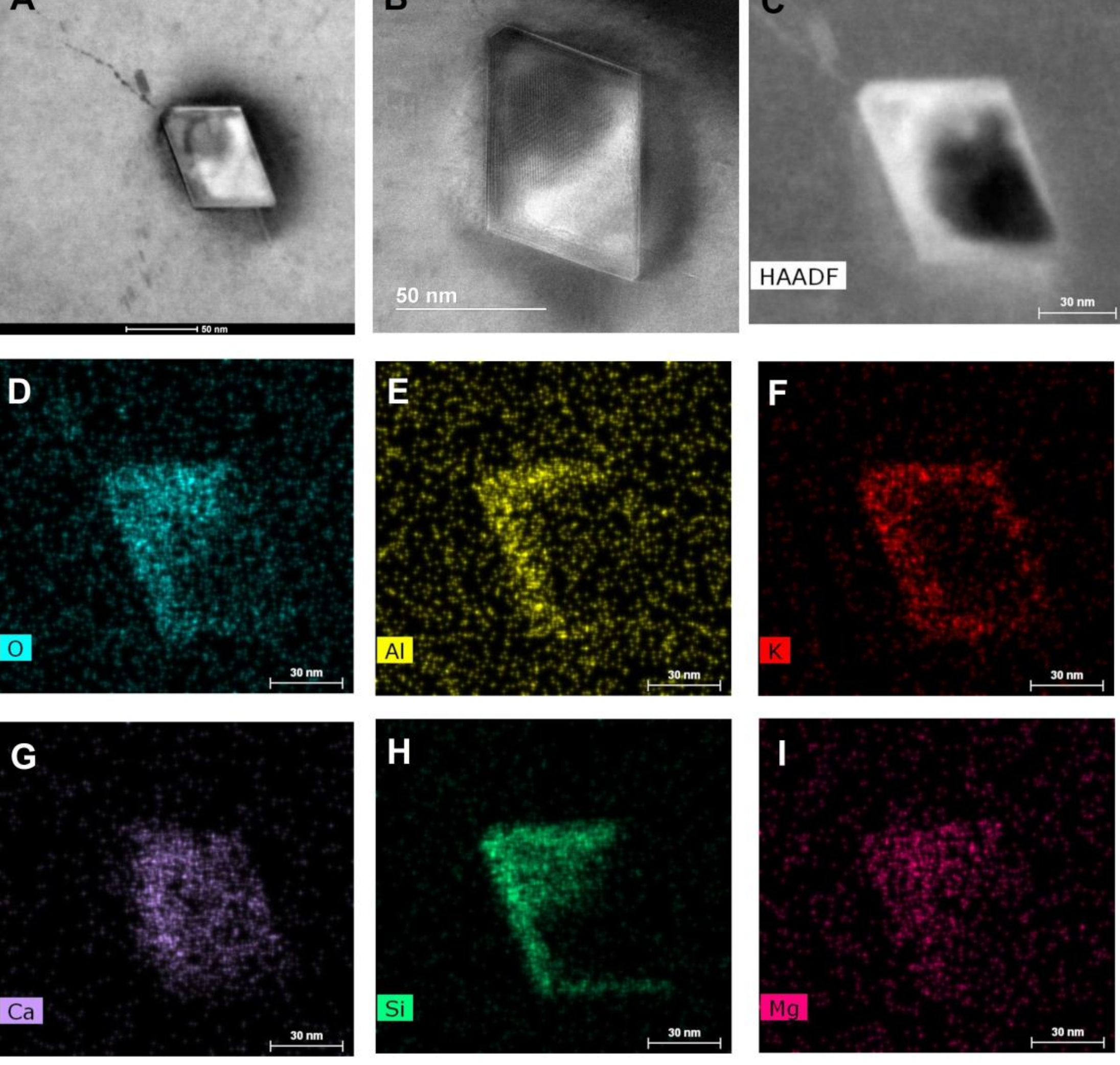

Fig. 5. Polyphase in U-14 diamond. A, B – bright field images. C – HAADF image; D-I – elemental maps.

*Dislocations*

In the zone with the Y-defects numerous perfect and partial dislocations in various configurations as well as dislocation loops are observed (Fig. 6). Important feature of this crystal zone is the presence of numerous clusters of point defects, generated by non-conservative dislocation motion (dislocation debris). These defects lack noticeable concentrations of impurities and, presumably, largely consist of vacancies and interstitials.

Although qualitatively the foil from peripheral part of the crystal is similar to that from the Y-defects zone (Fig. 7), density of dislocations and of defect clusters is, apparently, lower. At the same time, significant number of dislocation loops and dipoles is present. The smallest distance between individual dislocations in a dipole is ~11 nm, i.e. comparable with literature [Mussi et al., 2013]. Note that density of dislocations in foils from both zones of the specimen is markedly lower than in some brown diamonds where dislocation forest may occur [Laidlaw et al., 2021].

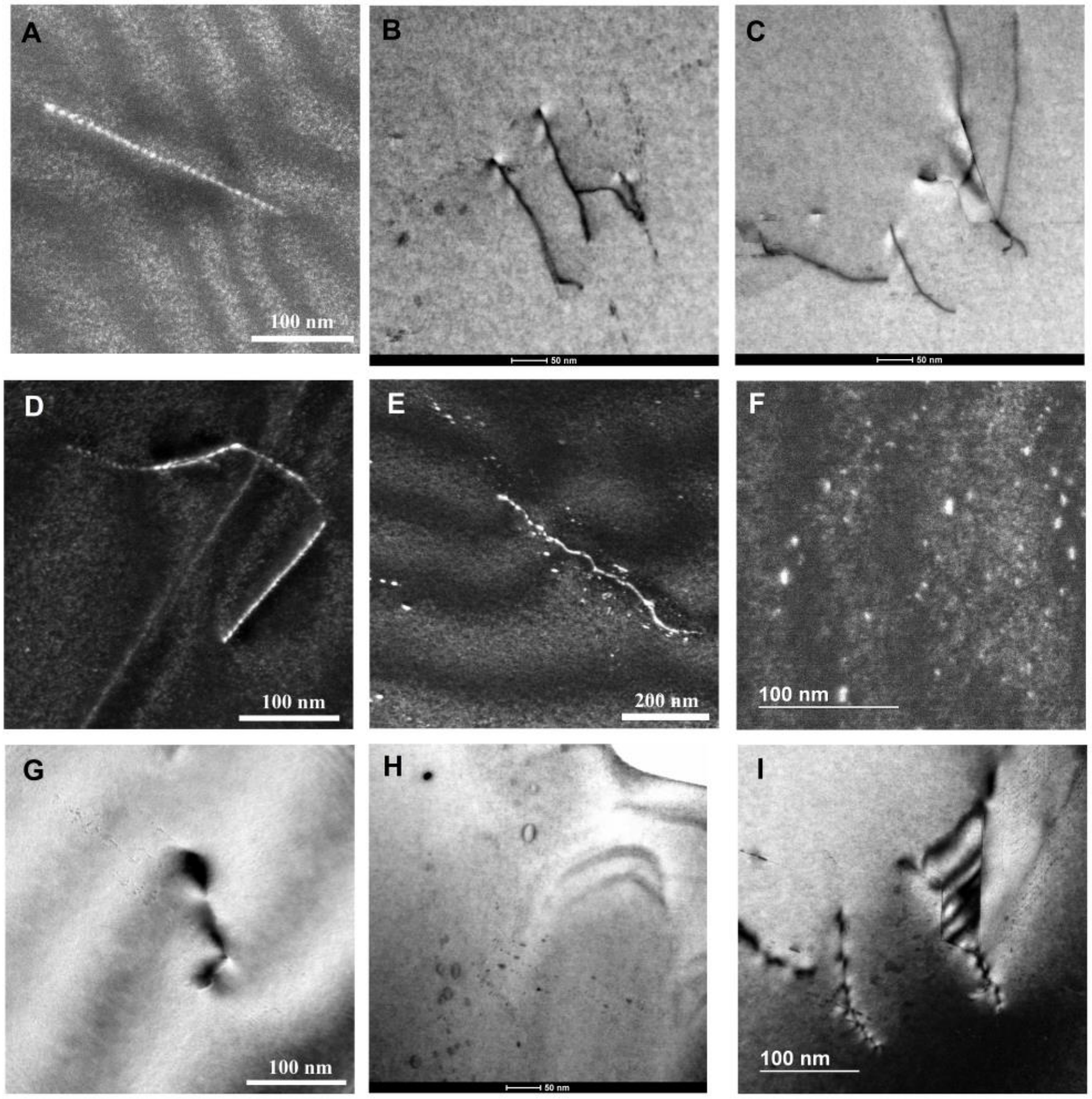

Fig. 6. Dislocations in a zone with Y-defects. A-E – various configurations of dislocations. F – dark-field image of a cluster of point defects. G – Polygonal dislocation loop and related chain of defect clusters. H – dislocation loops and clusters of defects. I – dislocation loops and clusters of defects generated by the dislocation movement. An object in the middle of the image I is not a stacking fault, the contrast is caused by an exposed inclusion.

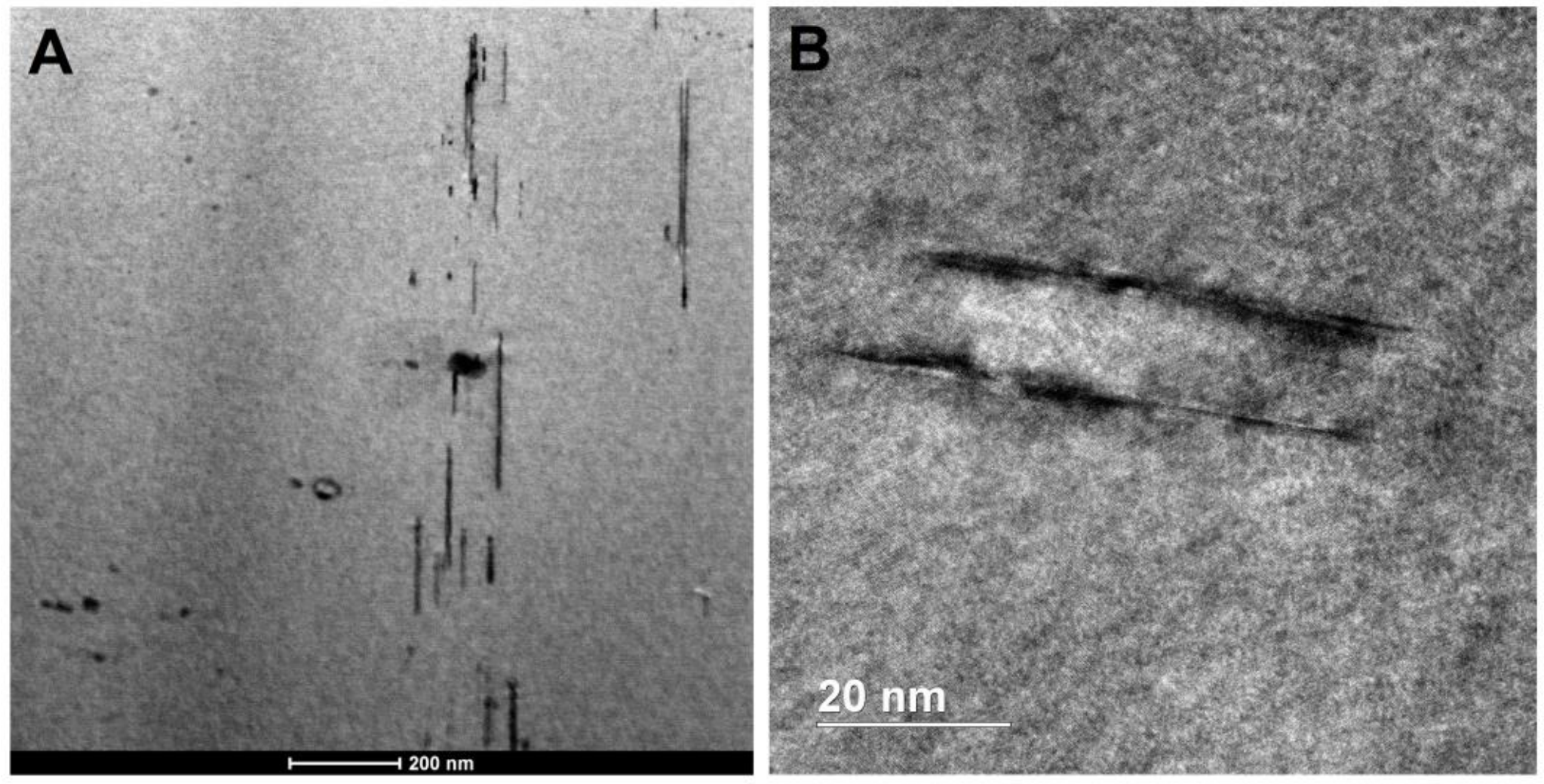

Fig. 7. Dislocations in peripheral zone of the crystal.

*Voids.*

Extended defects with morphology that could be approximated as rhombic plates are present in both zones of the studied diamond (Fig. 8). The largest dimension of these defects in the (111) plane reach 10-20 nm, but their thickness is not larger than 2-3 nm. These defects often form groups and may reside in intersecting {111} planes. Numerous attempts to detect impurities in these defects failed; we note that an idea that all the defects were opened during the sample preparation is clearly incorrect. The contrast variations in different imaging regimes point to low density of the entrapped material, thus, the defects could be described as agglomerations of vacancies.

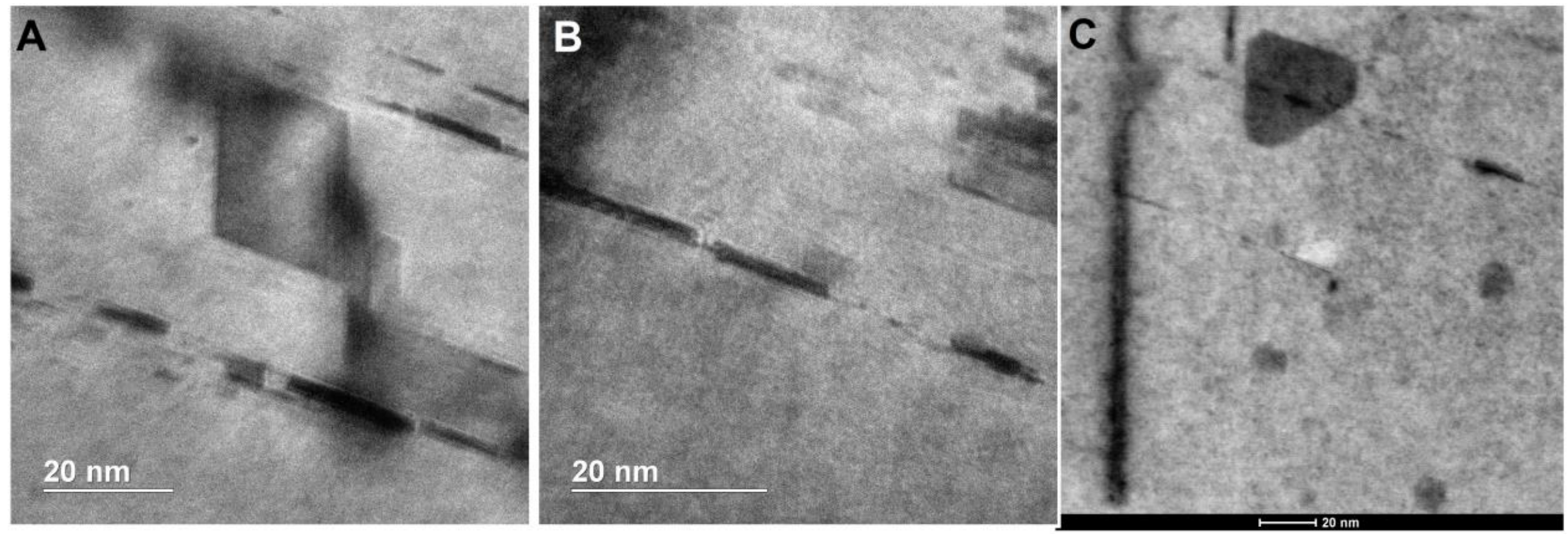

Fig. 8. A, B – «rhombic plates» in the diamond zone with Y-defects. C – defects in peripheral part of the crystal.

As indicated above, the rhombic plates are present both in the intermediate (with the Y-defects) and in peripheral parts of the crystal. However, in the peripheral part their number is, apparently, lower. However, in addition to the rhombic defects, in this zone another type of extended defects represented by weakly contrasted formations with distorted hexagonal shape is observed. These defects are very thin as suggested by imaging at tilt angles up to ±30°. In addition, in the peripheral part rare (quasi)hexagonal plates with sizes below 10 nm are encountered. The two later types of defects may represent graphitic flakes (e.g., Shiryaev et al., 2023).

## Discussion

*Spectroscopic manifestations of the Y-defects*

Absorption spectra of diamonds with the Y-defects always show weak systems in the ranges 1340–1400 and 2800-3500 $cm^{-1}$ [Day et al., 2025a,b; Hainschwang et al., 2012; 2020; Kupriyanov et al., 2020; Titkov et al., 2015]. Only the 1358 $cm^{-1}$ peak is unambiguously correlated with the Y-center [Titkov et al., 2015], all other bands are not directly related. Day and coauthors [2025a] suggest that the 1387 $cm^{-1}$ peak is also related to the Y-center; however, this assumption contradicts annealing experiments [Hainschwang, 2014], and to IR mapping of natural diamonds [Titkov et al., 2015].

Notably, in absolute majority of diamonds with the Y-defects, an IR absorption peak is recorded at diamond Raman frequency (1332 $cm^{-1}$); in such diamonds, it is usually ascribed to the X-defect ($N^+$). In ideal diamond lattice infra-red absorption at Raman frequency is forbidden by the selection rules, local distortions of the lattice symmetry by point defects may lift the limitation and corresponding features arise in spectra of many defects and even are manifested in Fano-resonance [Ekimov et al., 2022]. The correlation between the 1332 $cm^{-1}$ peak and maximum of the Y-center absorption (~1150 $cm^{-1}$) is not very convincing (this work and [Reutsky et al., 2017]), which is clearly explained by ambiguities in decomposition of the complex envelope. At the same time, absorptions at 1332 $cm^{-1}$ and 1358 $cm^{-1}$, the latter being related to the Y-defect, are directly correlated [Reutsky et al., 2017]. Note also that behavior of luminescing defects induced by electron irradiation and annealing in diamonds with the Y-centers (and with the 1332 $cm^{-1}$ peak) imply that the Y-center does not serve as an electron donor [Hainschwang, 2014], consequently, it is difficult to expect presence of the $N^+$ defects in such diamonds. Therefore, with high degree of confidence one may state that the Y-defect IR spectrum includes a sharp peak at diamond Raman frequency.

Based on the assumption that the Y-center manifests itself on the Raman frequency, we have attempted to find a correlation of corresponding IR absorption (after decomposition of the one-phonon region into defects-related components) and absorption between 2800-3500 $cm^{-1}$. A weak positive correlation is found between the 1332 and 3068 $cm^{-1}$ peaks. However, analysis of spectra from the [Titkov et al., 2015] paper does not confirm this correlation. This might indicate that the later line is caused by microinclusions and/or nanovoids (see TEM section) or be a mere coincidence.

On the profile of the defects distribution (Fig. 2C) one can see that total nitrogen concentration and Y-defects' absorption are inversely proportional; similar behavior was reported in [Hainschwang et al., 2012]. This dependency holds even despite uncertainties in evaluation of the A-defects contribution at 1282 $cm^{-1}$. These results contradict to model presented by [Day et al., 2025a], which assumes that the Y-defects represent an intermediate specie on the C→A transformation in diamonds with relatively high N content.

The Y-centers are encountered in two types of diamonds with somewhat different set of absorption lines with variable ratio of their intensities. The first type comprises diamonds with discrete absorptions, the most intense being 1353, 1358, 1374, 1387, 1432, 1445, 1465, 3050, 3144, 3154, 3189, 3311, 3343, 3394 $cm^{-1}$. Such spectra are encountered in diamonds with relatively high total nitrogen and high fraction of the A-defects [Day et al., 2025a, b]. These diamonds show a broad band peaked at ~1150 $cm^{-1}$, the bands at 1268, 1297 $cm^{-1}$ are absent. The second type is characterized by extended absorptions between 1340-1450 $cm^{-1}$ and 2700-3300 $cm^{-1}$ with numerous local maxima; the line at 2973 $cm^{-1}$ is the most intense. Such spectra are recorded in crystals (or in zones) with low concentrations of A- and C-defects; the absorption spectra show maxima at 1268 and 1297 $cm^{-1}$; the 1145 $cm^{-1}$ maximum is relatively narrow [Hainschwang et al., 2012, 2020; Kupriyanov et al., 2020]. Interestingly, graining is typical for natural type Ib diamonds and is not observed in crystals with prevailing Y-centers [Hainschwang et al., 2014]. Most likely, the relative fraction of diamonds of each type depends on diamondiferous province; reflecting conditions of formation and annealing.

*Extended defects*

Discovery of a novel type of extended defects in diamond is the most interesting result of the present study. In the crystal zone containing these defects, a set of narrow luminescence lines with intensities, changing with time (blinking). Despite obvious difficulties in establishing correlations between spectroscopic features and objects revealed by transmission electron microscopy, it seems that results obtained by different techniques are reconcilable with each other.

The nature of the planar rhombic defects and their formation mechanism remain unclear at present. It is known that octahedron is an equilibrium shape of a vacancy pore in silicon and germanium, which are isostructural with diamond [Voronkov, 1974; Mil'vidskii and Osvenskii, 1984]. In fact, on growth surface of high purity silicon octahedral (or truncated octahedron) pores produced by agglomeration of vacancies may be formed [Ueki et al., 1997]. In grains of polycrystalline diamond produced by pulse heating of graphite at high static pressure cubooctahedral voids with sizes from ~15 to 50 nm are encountered [Oshima et al., 1994]. Presumably, these pores represent clusters of non-equilibrium vacancies, formed during sample quench. Octahedral nitrogen-containing nanosized voidites serve as as final stage of aggregation of nitrogen defects in diamond [Kiflawi, Bruley, 2000]. In contrast with discussed above quasi-isometric pores in diamond and silicon, the defects observed in the present work are always strongly flattened. Their localization in intersecting planes and, presumably, random orientation relative to the growth surface prevents direct comparison with the growth-related nanopores in silicon.

On surfaces of some natural diamonds empty channels of macroscopic sizes (from several to hundreds microns) with rhombic opening [Lu et al., 2001, Schoor et al., 2016]; sometimes, the channels form zigzag pattern (e.g., Fig. 3 in [Lu et al., 2001]). Presumably, these defects represent Rose channels, which form at intersections of twin lamellae in course of brittle deformation of diamond [Schoor et al., 2016]; possibly, with subsequent etching [Lu et al. 2001]. Deformation (or mechanic) microtwinning of natural diamonds is well-known [Titkov et al., 2012]; in particular, formation of numerous elastic twins was reported [Gainutdinov et al., 2013]. Numerous intersecting slip planes were produced in experiments on uniaxial diamond compression at high hydrostatic pressures [Howell et al., 2012]. It was also shown that twin lamellae in rose diamonds consist of several parallel twinned domains with thickness of few nanometers [Gaillou et al., 2010]. Therefore, the defects observed by us might have been formed in course of brittle deformation of diamond with high density of intersecting twins, forming sets of Rose channels. The most important argument against this scenario is the lack of hints of twinning in electron diffraction patterns and the absence of a common plane, comprising all rhombic defects.

Nevertheless, the connection between the rhombic defects and slip planes is not fully excluded. According to ab initio calculations [Hounsome et al., 2006], at radius below 1.2 nm, i.e. for ≤ 200 vacancies, a vacancy disk is more favourable energetically than a dislocation loop. Clusters observed experimentally in brown diamonds few nanometers in size and, presumably, comprise relatively small number (~50) vacancies [Godfrey, Bangert, 2011]. Various mechanisms of dislocation dipole annihilation in diamond and silicon are considered employing different interatomic potentials in [Rabier, Pizzagali, 2010]. It is shown that if a separation of the dipole dislocations in (111) slip plane is less than a critical threshold, then planar vacancy clusters may form in (111) without creation of a strain field. According to calculations, the critical separation is 9.9 ± 4.4 nm [Rabier, Pizzagali, 2010]; this value is in good correspondence with experimental value of 7.4 ± 0.8 nm [Mussi et al., 2013] and with the present work. Linear dimensions of the voids found by us are in the 5-20 nm range. Therefore, one of possible scenario of their formation consists of annihilation of dislocation dipoles with subsequent increase of linear dimensions by consumption of vacancies, produced by non-conservative dislocation motion. However, the mechanism of stabilization and growth of these novel defects remains unknown.

Planar rhombic defects may be responsible for some of the peculiarities of the photoluminescence. The principal PL systems in the studied diamond comprise intense line at 904 nm (both PL and absorption), lines at 525, 689 and 698 nm were earlier reported for natural diamonds with C-defects [Hainschwang et al., 2006, 2013; Kupriyanov et al., 2020; Vasilev et al., 2020, 2024; Zudina et al., 2013]. Taking into account the novel nanosized voids, it is interesting to consider origin of narrow lines in the 800-900 nm range, characterized by chaotic variations in position and intensity of the lines at spots with minimal distances from each other. The broadening of the lines with temperature increase is typical for the luminescence rather than for Raman lines. Qualitatively similar photoluminescence behavior was reported for the lines in the range 500-550 nm in spectra of nanodiamonds synthesized from hydrocarbons at high static PT-conditions and assigned to recombination of donor-acceptor pairs on hydrogenated surfaces [Pasternak et al., 2024]. Analogous system was observed by one of the authors (E.A. Vasil'ev, pers. comm) in carbonado – nano-polycrystalline diamond aggregate. PL systems with numerous narrow lines were earlier reported for hydrogenated nanodiamonds [Kudryavtsev et al., 2018; Neliubov et al., 2023], and for other nanosized materials – Si [Schmitt et al., 2015], GaAs [Ha et al., 2015], $WSe_2$ [Koperski et al., 2015].

Presumably, the observed set of luminescence lines may be related to recombination of donor-acceptor pairs on walls of the nanosised defects. The differences in the lines' energies may be related to scatter in size of the nanovoids. Note that the PL lines mentioned above were recorded in nanodiamonds with hydrogenated surfaces. It is possible, that IR absorption bands between 2800-3300 $cm^{-1}$ typical for diamonds with the Y-defects may also be related to hydrogen decorating internal walls of the defects.

**Conclusions.**

Transmission Electron Microscopy of foils extracted from a Ib-IaA natural diamond (Yubileinaya kimberlite pipe) containing Y-defects shows that the sample underwent plastic deformation at high temperatures. Numerous dislocations, dislocation dipoles and loops are present. It is shown that the dislocation motion was non-conservative, leading to formation of clusters of point defects. Absence of noticeable concentrations of impurities imply that the clusters mostly consist of vacancies and interstitials.

Besides dislocations, extended defects with shapes that could be described as thin (1-3 nm) rhombic plates up to 5-20 nm in the largest dimension and lying in (intersecting) (111) planes. Analysis of contrast of the images and absence of impurities suggest that these objects represent nanosised voids (vacancy clusters). Presumably, formation of these defects is related to deformation of diamond. Several possible scenario of their formation is considered. Annihilation of dislocation dipoles with subsequent growth of dimensions due to vacancies generated by non-conservative dislocation motion is one possible mechanisms. Domination of this mechanism of diamond deformation instead of more common generation of slip planes, may be an important feature of diamonds with the Y-centers.

At excitation with 787 nm laser, in photoluminescence spectra of diamonds with the Y-centers numerous sharp lines are observed between 800-900 nm; their intensity and position varies irregularly at micron scale. Similar behavior is also established for other natural type Ib-IaA diamonds and carbonado. Chaotic changes of photoluminescence lines (blinking) was earlier reported for nanodiamonds synthesized from hydrocarbons and was assigned to donor-acceptor recombination on hydrogenated surfaces. We propose that unusual behavior of the luminescence lines in the studied crystal may be explained by recombination processes at walls of the observed nanovoids.

The Y-center proper and accompanying infra-red absorptions may be manifestations of various configurations of nitrogen and carbon interstitials [Day et al., 2025a; Hainschwang et al., 2012]. According to quantum chemistry calculations, a complex comprising four interstitials is stable in diamond [Goss et al., 2001]. For this defect IR-active vibrations are predicted to occur at 1332, 1355, 1389, 1722, 1741 $cm^{-1}$. Other configurations of interstitial atoms may also induce sharp IR absorption peaks between 1375 and 1925 $cm^{-1}$ [Cherati et al., 2025]. Thus, although at purely qualitative level, infra-red absorption spectra of a diamond zone with the Y-defects are consistent with the presence of abundant clusters of intrinsic defects, related to non-conservative motion of dislocations.

## REFERENCES

**Boyd S.R., Kiflawi I., Woods G.S.** The relationship between infrared absorption and the A defect concentration in diamond // Phil. Mag. B, 1994, v. 69, p.1149-1153. https://doi.org/10.1080/01418639408240185

**Cherati N.M., Hashemi A., Gali Á.** From Mono- to Hexa-Interstitials: Computational Insights into Carbon Defects in Diamond // arXiv:2512.06167, https://doi.org/10.48550/arXiv.2512.06167

**Clark C.D., Davey S.T.** One-phonon infrared absorption in diamond // J. Phys. C: Sol. St. Phys., 1984, 17, 1127–1140. https://doi.org/10.1088/0022-3719/17/6/020

**Collins A.T., Mohammed K.** Optical studies of vibronic bands in yellow luminescing natural diamonds // J. Phys. C: Sol. St. Phys. 1982, 15, 147–158. https://doi.org/10.1088/0022-3719/15/1/012

**Day M.C., Jollands M.C., Innocenzi F., Novella D., Nestola F., Pamato M.G.** Infrared spectroscopy of natural Type Ib diamond: insights into the formation of Y-centers and the early aggregation of nitrogen // Amer. Miner 2025a https://doi.org/10.2138/am-2024-9722

**Day M.C., Balan E., Caracas R., Jollands M.C., Innocenzi F., Novella D., Nestola F., Pamato M.G.** The first-formed hydrogen and nitrogen defects in natural diamond // Diam. Relat. Mater. 2025b, 158, p. 112654, https://doi.org/10.1016/j.diamond.2025.112654

**Ekimov E., Shiryaev A.A., Grigoriev Y., Averin A., Shagieva E., Stehlik S., Kondrin M.**, Size-Dependent Thermal Stability and Optical Properties of Ultra-Small Nanodiamonds Synthesized under High Pressure // Nanomaterials, 2022, 12(3), 351, https://www.mdpi.com/2079-4991/12/3/351/pdf

**Gaillou E., Post J.E., Bassim N.D., Zaitsev A.M., Rose T., Fries M.D., Stroud R.M., Steele A., Butler J.E.** Spectroscopic and microscopic characterizations of color lamellae in natural pink diamonds // Diamond Relat. Mater. 2010, 19, 1207–1220. https://doi.org/10.1016/j.diamond.2010.06.015

**Gainutdinov R.V., Shiryaev A.A., Boyko V.S., Fedourtchouk Y.** Extended defects in natural diamonds: an Atomic Force Microscopy investigation // Diamond and related materials, 2013, Vol. 40, 17-23. https://doi.org/10.1016/j.diamond.2013.09.006

**Gali A., Lowther J.E., Deák P.** Defect states of substitutional oxygen in diamond // J. Phys.: Condens. Matter 2001, 13, 11607. DOI 10.1088/0953-8984/13/50/319

**Godfrey I.S., Bangert U.** An analysis of vacancy clusters and sp2 bonding in natural type IIa diamond using aberration corrected STEM and EELS // J. Phys.: Conf. Ser. 2011, 281, 012024. https://doi.org/10.1088/1742-6596/281/1/012024

**Goss J.P., Coomer B.J., Jones R., Shaw T.D., Briddon P.R., Oberg S.** Interstitial aggregates in diamond // Diamond and Related Materials, 2001, 10, 434-438.

**Ha N., Mano T., Chou Y-L., Wu Y-N., Cheng S-J., Bocquel J., Koenraad P. M., Ohtake A., Sakuma Y., Sakoda K., Kuroda T.** Size-dependent line broadening in the emission spectra of single GaAs quantum dots: impact of surface charge on spectral diffusion // Physical Review B, 2015, v. 92(7), 075306-1/8. Doi:10.1103/PhysRevB.92.075306

**Hainschwang T,** Diamants de type Ib: Relations entre les propriétés physiques et gemmologiques des diamants contenant de l’azote isolé // These de doctorat, 2014, Universite de Nantes

**Hainschwang T, Fritsch E., Notari F., Rondeau B.** A new defect center in type Ib diamond inducing one phonon infrared absorption: The Y center //Diamond Relat. Mater, 2012, v. 21. p. 120-126. https://doi.org/10.1016/J.DIAMOND.2011.11.002

**Hainschwang T., Notari F., Fritsch E., Massi L.** Natural, untreated diamonds showing the A, B and C infrared absorptions (“ABC diamonds”), and the H2 absorption // Diamond Relat. Mater, 2006, v. 15, p. 1555-1564. https://doi.org/10.1016/j.diamond.2013.07.007

**Hainschwang T., Notari F., Pamies G.** A Defect Study and Classification of Brown Diamonds with Non-Deformation-Related Color // Minerals, 2020, 10, 914. https://doi.org/10.3390/min10100914

**Howell D., Piazolo S., Dobson D.P., Wood I.G., Jones A.P., Walte N., Frost D.J., Fisher D., Griffin W.L.** Quantitative characterization of plastic deformation of single diamond crystals: A high pressure high temperature (HPHT) experimental deformation study combined with electron backscatter diffraction (EBSD) // Diamond & Related Materials, 2012, 30, 20–30. https://doi.org/10.1016/j.diamond.2005.12.029

**Hounsome L.S., Jones R., Martineau P.M., Fisher D., Shaw M.J., Briddon P.R., Öberg S.** Origin of brown coloration in diamond // Physical Review B, 2006, 73, 125203. https://doi.org/10.1103/physrevb.73.125203

**Kiflawi I., Bruley J.** The nitrogen aggregation sequence and the formation of voidites in diamond // Diamond and Relat. Mater, 2000, v. 9 (1), p. 87—93. Doi:10.1016/S0925-9635(99)00265-4

**Koperski M., Nogajewski K., Arora A.V., Cherkez, Mallet P., Veuillen J.-Y., Marcus J., Kossacki P., Potemski M.** Single photon emitters in exfoliated $WSe_2$ structures // Nature Nanotech, 2015, v. 10, p. 503–506. Doi:10.1038/nnano.2015.67

**Kovalchuk O.E.** Genetic interpretation of investigation of diamonds by infra-red spectroscopy // Abstracts of Ural mineralogical school. 2024. pp. 95-96

**Kudryavtsev O.S., Ekimov E.A., Romshin A.M., Pasternak D.G., Vlasov I.I.** Structure and Luminescence Properties of Nanonodiamonds Produced from Adamantane // Phys. Status Solidi A Appl. Mater. Sci. 2018, 215, 1800252. https://doi.org/10.1002/pssa.201800252

**Kupriyanov I.N., Palyanov Y.N., Kalinin A.A., Shatsky V.S.** Effect of HPHT Treatment on Spectroscopic Features of Natural Type Ib-IaA Diamonds Containing Y Centers // Crystals 2020, 10, 378; doi:10.3390/cryst10050378

**Lai M.Y, Hardman M.F., Eaton-Magaña S., Breeding C.M., Schwartz V.A., Collins A.T.,** Spectroscopic characterization of diamonds colored by the 480 nm absorption band // Diamond and Related Materials, 2024, 142, 110825. https://doi.org/10.1016/j.diamond.2024.110825

**Laidlaw F.H.J., Diggle P.L., Breeze B.G., Dale M.W., Fisher D., Beanland R.** Spatial distribution of defects in a plastically deformed natural brown diamond // Diamond and Related Materials, 2021, 117, 108465. https://doi.org/10.1016/j.diamond.2021.108465

**Lawson C.S., Fisher D., Hunt D.C., Newton M.E.** On the existence of positively charged single-substitutional nitrogen in diamond // J Phys Condens Matter, 1998, v. 10, p. 6171-6180. https://doi.org/10.1088/0953-8984/10/27/016

**Lu T., Shigley J.E., Koivula J.I., Reinitz I.M.** Observation of etch channels in several natural diamonds // Diamond and Related Materials 2001, 10, 68-75. https://doi.org/10.1016/S0925-9635(00)00407-6

**Malogolovets V.G., Nikityuk N.I.,** Vibrational frequencies of diamond lattice induced by substitutional impurities // Superhard. Mater. 1981, 5, 28–31

**Mil'vidskii M.G., Osvenskii V.B.,** Structural defects in semiconductors single crystals, Moscow, Metallurgy, 1984, 256 p.

**Mussi A., Eyidi D., Shiryaev A.A., Rabier J.** TEM Observations of dislocations in plastically deformed diamond // Physica Status Solidi A, 2013, 210(1), 191–194. https://doi.org/10.1002/pssa.201200483

**Nadolinny V.A., Palyanov Yu.N., Rakhmanova M.I., Borzdov Yu.M., Komarovskikh A.Yu., Shatsky V.S., Ragozin A.L., Yurjeva O.P.** Investigation of the Y centers in cubic plastically deformed type Ib diamonds (Yakutia placers) // Diamond and related materials, 2025, v. 151, 111821. https://doi.org/10.1016/j.diamond.2024.111821

**Neliubov A.Y., Eremchev I.Yu., Drachev V.P., Kosolobov S.S., Ekimov E.A., Arzhanov A.I., Tarasevich A.O., Naumov A.V.** Enigmatic color centers in microdiamonds with bright, stable, and narrow-band fluorescence // Phys. Rev. B, 2023, 107, L081406. https://doi.org/10.1103/physrevb.107.l081406

**Oshima R., Togaya M., Hattori K.** Vacancy-Type Defect Clusters in Diamond Grown from Molten Graphite at High Pressures // Jpn. J. Appl. Phys. 1994, v. 33, L440-L442

**Palyanov Y.N., Kupriyanov I.N., Sokol A.G., Borzdov Y.M., Khokhryakov A.F.**, Effect of $CO_2$ on crystallization and properties of diamond from ultra-alkaline carbonate melt // Lithos, 2016, 265, 339–350, https://doi.org/10.1016/j.lithos.2016.05.021.

**Pasternak D.G., Romshin A.M., Bagramov R.H., Galimov A.I., Toropov A.A., Kalashnikov D.A., Leong V., Satanin A.M., Kudryavtsev O.S., Gritsienko A.V., Chernev A.L., Filonenko V.P., Vlasov I.I.** Donor–Acceptor Recombination Emission in Hydrogen-Terminated Nanodiamond // Adv. Quantum Technol. 2024, 2400263, DOI: 10.1002/qute.202400263

**Rabier J., Pizzagalli L.** Dislocation dipole annihilation in diamond and silicon // J. Phys.: Conf. Ser. 2011, 281 012025. https//doi.org/10.1088/1742-6596/281/1/012025

**Reutsky V.N., Shiryaev A.A., Titkov S.V., Wiedenbeck M., Zudina N.N.** Evidence for Large Scale Fractionation of Carbon Isotopes and of Nitrogen Impurity during Crystallization of Gem Quality Cubic Diamonds from Placers of North Yakutia // Geochemistry International, 2017, 55(11), 988–999

**Schoor M., Boulliard J.C., Gaillou E., Hardouin Duparc O., Estève I., Baptiste B., Rondeau B., and Fritsch E.** Plastic deformation in natural diamonds: Rose channels associated to mechanical twinning // Diamond and related materials, 2016, v. 66, 102-106. https//doi.org/10.1016/j.diamond.2016.04.004

**Schmitt S., Sarau G., Christiansen S.** Observation of strongly enhanced photoluminescence from inverted cone-shaped silicon nanostructures.// Sci Rep, 2015, v. 5, 17089, Doi: 10.1038/srep17089

**Shiryaev A.A., Chesnokov Y., Vasiliev A.L., Hainschwang T.** Exsolution of oxygen impurity from diamond lattice and formation of pressurized $CO_2$-I precipitates // Carbon Trends, 2023, v. 11, 100270. https://doi.org/10.1016/j.cartre.2023.100270

**Titkov S.V., Krivovichev S.V., Organova N.I.** Plastic deformation of natural diamonds by twinning: evidence from X-ray diffraction studies // Mineral. Mag. 2012, v.76, 143-149. https://doi.org/10.1180/minmag.2012.076.1.143

**Titkov S.V., Shiryaev A.A., Zudina N.N., Zudin N.G., Solodova Yu.P.** Defects in cubic diamonds from the placers in the northeastern Siberian platform: results of IR microspectrometry // Russian Geology and Geophysics, 2015, V. 56, 354-362

**Ueki T., Itsumi M., Takeda T.** Octahedral Void Structure Observed in Grown-In Defects in the Bulk of Standard Czochralski-Si for MOS LSIs // Jpn. J. Appl. Phys. 1997, v. 36, 1781-1785

**Vasilev E.A., Zedgenizov D.A., Klepikov I.V.** The enigma of cuboid diamonds: the causes of inverse distribution of optical centers within the growth zones // Journal of Geosciences, 2020, v. 65(1), p. 59 – 70, Doi:10.3190/jgeosci.301

**Vasilev E., Gubanov N., Zedgenizov D.** Point defects in coated diamonds // Diam. Relat. Mater., 2024, v. 148, p. 111519. https://doi.org/10.1016/j.diamond.2024.111519

**Voronkov V. V.,** Formation of vacancy pores during the cooling of germanium and silicon // Sov. Phys. Crystallogr. 1974, 19, 137

**Woods G.S., Collins A.T.** Infrared absorption spectra of hydrogen complexes in type I diamond // J. Phys. Chem. Sol. 1983, v. 44 (5), 471–475.

**Zudina, N.N., Titkov, S.V., Sergeev, A.M., Zudin, N.G.,** Peculiarities of photoluminescence centers in cubic placer diamonds of different colors from the northeastern Siberian Platform. // Zapiski RMO, 2013, CXLII (4), 57–72.